\documentclass[12pt, letterpaper]{article}
\usepackage[top=1in, bottom=1.5in, left=1in, right=1in]{geometry}
\usepackage[utf8]{inputenc}
\usepackage{booktabs}
\usepackage{hyperref}
\usepackage{amssymb}
\usepackage{graphicx}

\title{A near-Sun Solar System Twilight Survey with LSST}
\author{Rob Seaman, Paul Abell, Eric Christensen, Michael S.~P.\@ Kelley, \\ Megan E. Schwamb, Renu Malhotra, Mario Juri\'c, Quanzhi Ye \\ Michael Mommert, Matthew M. Knight, Colin Snodgrass, Andrew S. Rivkin}
\date{November 30, 2018}

\begin{document}

\maketitle

\begin{abstract}
We propose a LSST Solar System near-Sun Survey, to be implemented during twilight hours, that extends the seasonal reach of LSST to its maximum as fresh sky is uncovered at about 50 square degrees per night (1500 sq. deg. per lunation) in the morning eastern sky, and surveyable sky is lost at the same rate to the western evening sky due to the Earth’s synodic motion. By establishing near-horizon fence post picket lines to the far west and far east we address Solar System science use cases (including Near Earth Objects, Interior Earth Objects, Potentially Hazardous Asteroids, Earth Trojans, near-Sun asteroids, sun-grazing comets, and dormant comets) as well as provide the first look and last look that LSST will have at the transient and variable objects within each survey field. This proposed near-Sun Survey will also maximize the overlap with the field of regard of the proposed NEOCam spacecraft that will be stationed at the Earth's L1 Lagrange point and survey near quadrature with the Sun. This will allow LSST to incidently follow-up NEOCam targets and vice-versa (as well as targets from missions such as Euclid), and will roughly correspond to the Earth's L4 and L5 regions.

\end{abstract}

\section{White Paper Information}

\textbf{Corresponding Author}: Rob Seaman (seaman@lpl.arizona.edu)

% other authors: Paul Abell, Eric Christensen, Cesar Fuentes, Robert Jedicke, Mario Juri\'c, Michael S.~P.\@ Kelley, Matthew Knight, Hsing-Wen Lin, Renu Malhotra, Michael Mommert, Andy Rivkin, Meg Schwamb, Quanzhi Ye

\begin{enumerate} 
\item {\bf Science Category:} Taking an Inventory of the Solar System
\item {\bf Survey Type Category:}  Other - twilight survey 
\item {\bf Observing Strategy Category:} an integrated program with science that hinges on the combination of pointing and detailed observing strategy
\end{enumerate}  

\clearpage

\section{Scientific Motivation}
We propose a twilight Solar System survey with the Large Synoptic Survey Telescope (LSST) via near-horizon fencepost picket lines to the far west and far east. This proposed survey would address Solar System science objectives described below that are not achievable with the Wide-Fast-Deep (WFD) survey as described in \cite{Ivezic}.  Twilight observations of Solar System  bodies at low solar elongations will significantly increase the solar phase angle coverage of targets discovered during WFD operations, improving photometric models and promoting detection of likely activity in a range of Solar System small bodies. Indeed, some objects with orbits very similar to or interior to the Earth's (such as Earth Trojans or Atira asteroids) may otherwise never enter the LSST field of regard. A near-Sun twilight survey will also enable observations of bodies much closer to the Sun than the WFD Survey, significantly increasing the chances for detecting activity in these objects. LSST's large aperture and efficient operations can open a new window into the Solar System by working near-Sun.
\newline
\\
\textbf{Earth Trojan asteroids:} Earth Trojan asteroids, if they exist, would be librating at the Earth-Sun L4 and L5 Lagrange points. The population of Earth Trojans is presently not well constrained, representing a major gap in the inventory of small bodies in near-Earth space \cite{Malhotra:2018}. Numerical simulations find that asteroids in low-eccentricity ($e\lesssim$ 0.1), low inclination ($i\lesssim12^\circ$) orbits in Earth's Trojan and horseshoe regions would be only weakly chaotic on gigayear timescales, so asteroids can stably orbit near Earth's L4 and L5 regions for timescales comparable to the age of Earth \cite{Tabachnik:2000, Malhotra:2011, Cuk:2012}. Previous observational surveys of Earth's Trojan regions have derived an upper limit on their number density of 3 per sq-deg brighter than R=22.8 \cite{Whiteley:1998, Cambioni:2018}. This upper limit is equivalent to a population of several hundred Earth Trojans larger than a few hundred meters in size \cite{Wiegert:2000}. Presently, only one Earth Trojan is known, 2010 TK7, a $\sim$~{300m} size object that was discovered by WISE \cite{Connors:2011}. Improving the observational census of Earth’s Trojans is important for improving knowledge of the sources of planetary impactors for contemporary as well as historical impact cratering of the Earth and the Moon. Additionally, Earth Trojans would have low relative velocity with Earth, making them attractive targets for spacecraft missions. Discovery of a primordial population of Earth’s Trojans would provide direct access to the building blocks of our home planet. \newline
\\
\textbf{Near-Earth asteroids (NEAs):} NEAs are a diverse population, only some of which will be observed in LSST's WFD field of regard. Potentially Hazardous Asteroids (PHAs) are known to present as an overabundance in observational sweet spots \cite{Chesley:2004} near-Sun and near the ecliptic, overlapping the Earth's L4 and L5 regions. LSST's large aperture will compensate for the unfavorable phase angle that makes sweet spot observations difficult for smaller survey telescopes. Interior Earth Objects (IEOs), also known as Atira asteroids, are constrained to low solar elongations, thus being rare discoveries for current NEA surveys. LSST could increase the known IEO catalog dramatically, with a near-Sun twilight survey. \newline
\\
\textbf{near-Sun asteroids:} near-Sun asteroids have close encounters with the Sun due to small perihelion distances. Granvik et al.~\cite{Granvik2016} found a lack of near-Sun asteroids in the NEA population, which they attribute to the catastrophic disruption of these objects through thermal fracture. An example for repeated comet-like activity has been  found in near-Sun asteroid (3200) Phaethon \cite{Jewitt2010, Jewitt2013, Li2013}, supporting the idea of thermal fracture driving the activity due to its proximity to the Sun (q = 0.14~au). Studying near-Sun asteroids provides insights into the exact activation mechanisms, as well as asteroid properties under extreme conditions. \newline
\\
\textbf{Sun-grazing comets} Sun-grazing comets are observed by space-based solar observatories.  These objects have photometric lightcurves suggesting a cometary nature (i.e., sublimation induced mass-loss, dust scattering in an unresolved coma), and are occasionally bright enough to display tails or comae, securing this identification.  Despite their significant numbers, such comets are generally difficult to detect with ground-based assets \cite{Knight2010, Ye2014}.  This is primarily due to their small sizes ($\lesssim$50-m radius \cite{Knight2010}), orbital uncertainties (solar-observing telescopes have large pixel scales and short observing arcs), and the fact that the smallest objects are destroyed near perihelion.  In addition, their physical responses to insolation may be very different from the typical comet due to previous passages through the extreme near-Sun environment.  Does this processing create a thick insulating layer, quenching activity at larger distances?  With LSST discoveries and follow-up studies of sun-grazing comets at 1~au, we can finally address this question. The Kreutz sun-grazer family is the most likely to be observed near 1 au, based on their orbits and high numbers (1 comet near perihelion every 2 days \cite{Knight2010}).  Given their faintness ($\gtrsim$22 mag \cite{Ye2014}), low sky density ($<$0.1 deg$^{-2}$ \cite{Ye2014}), and low solar elongations (40--50 deg when near 1~au), an LSST twilight survey will have the best chance to discover a Kreutz sungrazing comet at 1~au before it is destroyed by the Sun near perihelion.  However, non-detections by LSST will have statistically significant consequences on our understanding of the population's surface volatiles.  Observing such comets under more typical circumstances will help connect this population to the population of comets as a whole, address how insolation processes cometary surfaces, and consider how small active comets can be. \newline
\\
\textbf{Dormant comets: }Dormant comets are asteroids of cometary origin still able to harbor volatiles in their sub-surface layers \cite{Wetherill1991, Mommert2015}. While most dormant comets are hiding in the asteroid populations, cases have been found in which dormant comets show faint activity, including near-Earth object (3552) Don Quixote \cite{Mommert2014}. The characterization of the volatile content of this potentially volatile-rich population of asteroids provides important constraints on the origin of terrestrial water, as well as the life cycle of comets (see above). Mommert et al.~\cite{Mommert2015} found that  $\sim$10\% of near-Earth asteroids $\gtrsim$ 1~km in diameter are of cometary origin and may still harbor volatiles. The only way to find proof of their existence is the direct detection of cometary activity in those bodies via imaging of coma/tails or photometric brightening, which requires observations of these objects close to perihelion, and hence at low solar elongations. The WFD Survey will not observe at low elongation. Only a dedicated twilight survey can reveal the true nature of potential dormant comets.

\section{Technical Description}

\subsection{High-level description}
\textbf{We propose to utilize available twilight time to observe fields within $\sim${40} degrees of the ecliptic at low solar elongations and high phase angles, for near-Sun Solar System bodies.}  

The great majority of observations made with the Large Synoptic Survey Telescope will be in service of some variation of the WFD “universal” cadence, chosen as a consensus among LSST’s diverse science drivers. This has three limitations from the point of view of Solar System observations: 1) the maximum airmass will be 1.5, and perhaps even less depending on the final selected cadence, 2) only rarely will there be more than 2 visits per night to any field, and 3) filter selection will be complex and not optimized for Solar System objects.

Airmass 1.5 corresponds to an elevation of 42 degrees, meaning that during WFD operations the telescope will approach no closer than this to the horizon. It is well known that from LSST’s site 30 degrees south of the equator this will truncate coverage of the northern ecliptic region which will require a special mini-survey. But access to the eastern and western horizons will be similarly constrained; indeed one can anticipate that most LSST observations will be made near the meridian at large solar elongations and low phase angles. Solar opposition should be well covered by the WFD portion of the survey, at least during the southern winter months, but access to low solar elongations will require special handling. Solar System objects occupying the near-Sun region are only accessible at low elevations soon after sunset or before sunrise. A twilight survey is one way to address the science use cases described in this document without impacting WFD or other mini-survey operations.

The limitation of the WFD survey to 2 visits per field per night necessitates the multi-tracklet linking concept of the Moving Object Processing System (MOPS)]\footnote{See LSE-30 (\url{http://ls.st/LSE-30}) and  LDM-156 (\url{http://ls.st/LDM-156)}}. A moving object detection will only be confirmed after at least three candidate tracklets are successfully linked across multiple nights. Individual 2-exposure tracklets will often be false. By contrast, ongoing Near Earth Object (NEO) surveys provide high-purity same-night detections through motion inferred directly from short-arc multiple-exposure tracklets. Since the WFD only revisits a field several nights later, it will take more than a week (significantly more depending on weather, moon, etc) to confirm most initial detections. This is unwieldy near-Sun since sky is vanishing behind the sun in the western sky at the rate of 7 degrees per week (two LSST Field-of-view widths) and unsurveyed sky is appearing in the east at the same rate.

Two alternatives exist for near-Sun fields: 1) a twilight NEO survey could revisit the same fields on successive nights, shortening the discovery interval to a minimum of three nights, or 2) an NEO-rate-optimized cadence of 3-4 exposures in the available 20-30 twilight minutes would produce NEO candidates for same-night submission to the Minor Planet Center’s NEO Confirmation Page\footnote{\url{https://www.minorplanetcenter.net/iau/NEO/toconfirm_tabular.html}} as with current surveys. The second option is preferred since more sky can be covered without repeated fields, and in either event such discoveries will require external follow-up to lengthen the orbital arc. The first option would also make it difficult for the Solar System twilight near-Sun survey to be scheduled in alternation with other twilight surveys or to allow for poor weather.

In either case several astrometric follow-up sites exist to the west of Cerro Pachon in both the southern and northern hemisphere. Since twilight operations, even for LSST, means a brighter limiting magnitude, targets will be accessible for follow-up by relatively modest aperture facilities either through longer "track and stack" observations, or simply by waiting for the darker night sky.

The 3.5 degree LSST field width corresponds well to even a one-field fencepost pattern as the sky shifts about a degree per day, thus traversing the LSST field of view in the requisite three nights. Morning near-Sun objects will generally move into the nightly WFD footprint during the same or future lunations. By contrast, evening twilight is a celestial subduction zone; LSST will not get another look at evening near-Sun discoveries during the same orbital apparition. Thus by observing near-Sun the seasonal reach of LSST is extended to its maximum for transient targets of all kinds.

The short twilight time interval of 20-30 minutes places a premium on efficient operations. Slews should be limited to adjoining fields with a single (or a few) larger jumps back to the beginning of the nightly raster. Similarly filter changes should be minimized, ideally to a single twilight filter choice each night, presumably chosen to be the first filter to be used during WFD operations in the evening or last filter during the morning.

\subsection{Footprint -- pointings, regions and/or constraints}

\begin{figure}[hbtp]
\centering
\includegraphics[width=16cm]{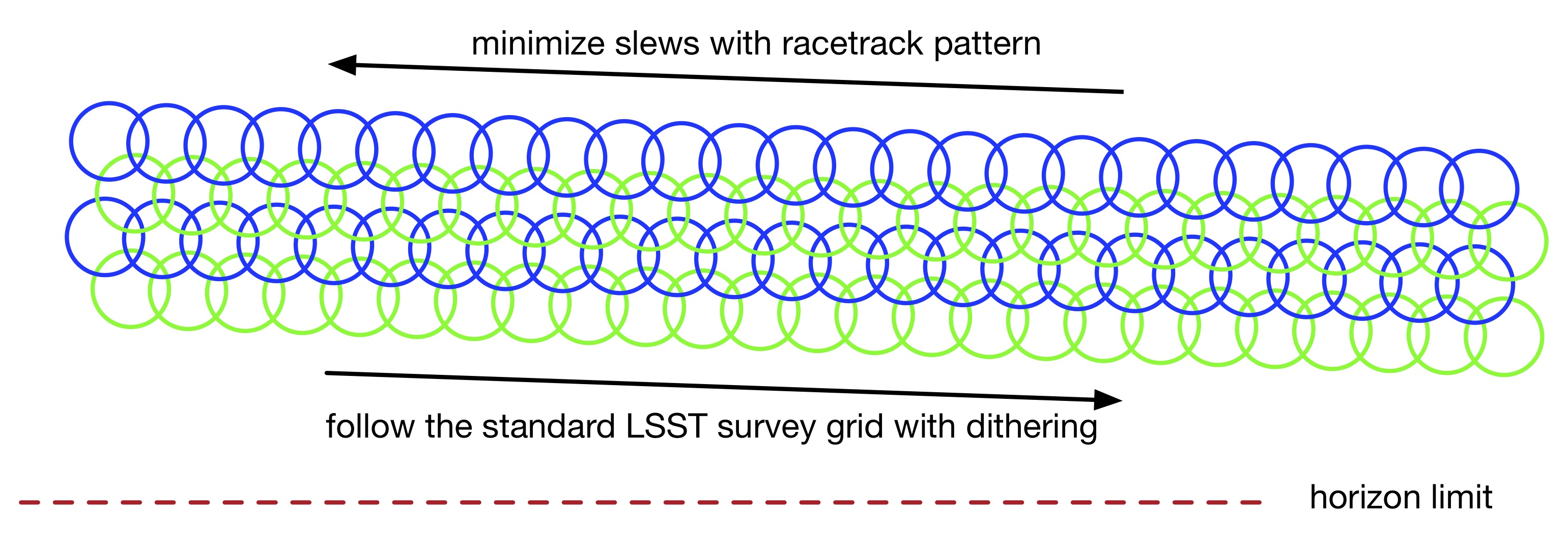}
\caption{Sketch of maximum sky coverable during 20 minute twilight with 1 second exposures and three image repeats separated by 400 seconds (assuming 2s readout + 1s shutter overhead with no extra step \& settle cost). Totals 100 fields for about 800 sq. degrees.}
\label{fig:tiling}
\end{figure}

Our proposed footprint is a picket line ranging as far west in the evening sky and as far east in the morning sky as is practical for LSST. Our aim is to effectively use twilight time that would otherwise not be used. Any sky coverage within 40 degrees of the ecliptic (perhaps more to the south) is useful to search for the low solar elongation and near-Sun small body populations. Figure 1 illustrates our recommended arrangement corresponding to three repeats of the shortest possible exposures, 1 second. With overlap, the pattern would reach approximately 40 degrees on either side of the ecliptic and cover a swath of about 10 degrees, but our footprint can easily use the main LSST field tessellation.

\begin{figure}[hbtp]
\centering
\includegraphics[width=12cm]{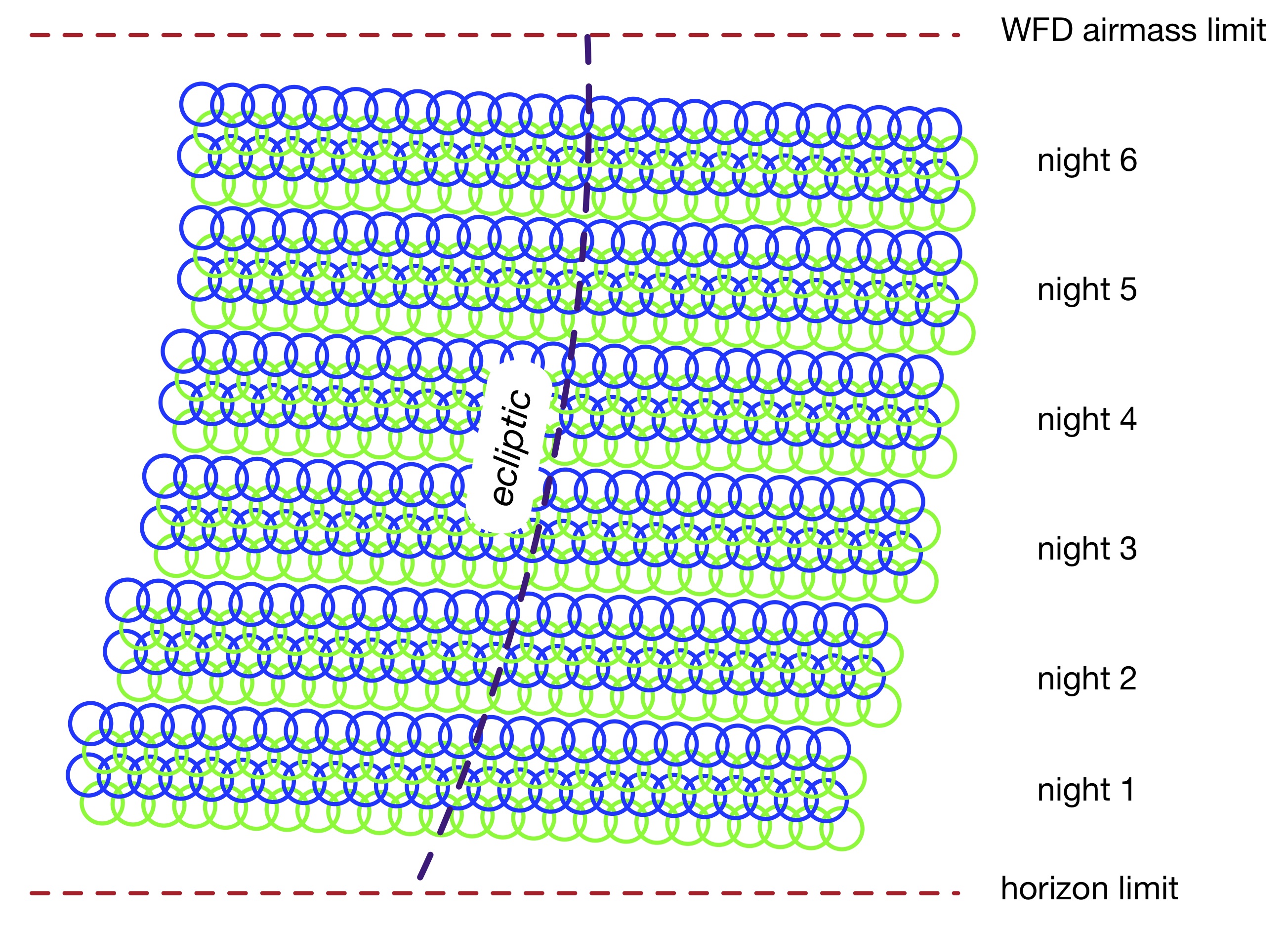}
\caption{Sketch of L4 (morning) or L5 (evening) tiling for maximum nightly sky coverage, not to scale. If twilight allocation is shorter, exposures longer, or number of revisits is greater than three the nightly coverage will decrease and more nights will be needed to complete tiling. West (evening) tiling would be inverted from morning tiling, marching from WFD limit toward horizon. As shown, pattern will move north and south with ecliptic. Precise tiling grid will be adopted from WFD survey and will not closely resemble sketch.}
\label{fig:racetrack}
\end{figure}

Even if the east/west footprints remained stationary in elevation from night to night, the sky would scroll behind at a rate of roughly one degree per night. Rather, this proposal would tile the area from the telescope's horizon limit up toward the elevation limit of the WFD survey so as to cover the available field of regard of the Earth's L4 and L5 Lagrange points over the course of months, with the surveying rate and opportunity to repeat as permitted by other twilight programs and weather. This would also extend the seasonal reach of LSST for patches of sky beyond the WFD field of regard for transients of all kinds.

\subsection{Image quality}

While Exposure targets during the near-Sun twilight survey will benefit from observations made in excellent conditions, normal or even poor seeing will still enable discovery of brighter objects. Thus, we have no specific image quality request. 

\subsection{Individual image depth and/or sky brightness}

No constraints. Moving object surveys, in particular during twilight, are necessarily serendipitous. Exposure time may be adjusted to compensate for the varying sky brightness during twilight, but our proposed survey aims for 1s exposures when possible. We refer the reader to the Community Observing Strategy Evaluation Paper (COSEP)\cite{COSEP} which describes the minimum expected exposures as a function of time in twilight. 

\subsection{Co-added image depth and/or total number of visits}

No constraints. Solar System bodies would be found in individual exposures, thus we have no constraints or restrictions on co-added depth.  Repeat visits from year-to-year would allow LSST to follow-up its own discoveries such as Earth Trojans (and to make additional discoveries), but once a minimum orbital arc is acquired, other telescopes should be able to extend the arcs through long duration exposures even with relatively small apertures.

\subsection{Number of visits within a night}

Given the short duration of each twilight period, a premium will be placed on generating tracklets that are actionable in near real-time. This will require 3 or more visits for each field within each twilight with candidate NEOs and other known and unknown object astrometry published to the Minor Planet Center (MPC) in the normal fashion. (A 4 or 5 repeat cadence would minimize the loss of moving objects in chip gaps, but would correspondingly reduce the nightly twilight area coverage.) Our proposed cadence is taking one visit before slewing to the neighboring field, but with a racetrack-style field revisit strategy of at least three visits per twilight. Future Operations Simulator (OpSim) \cite{opsim} simulations would be useful to evaluate this strategy (for twilight fields only) against the WFD MOPS strategy of multiple tracklet linking, but a 3 visit cadence is well-known to work efficiently for ongoing NEO surveys.

\subsection{Distribution of visits over time}

If a three-or-more nightly visit cadence is used to produce near-real-time actionable tracklets, night-by-night visit entanglements will be minimized. A two-nightly-visit cadence of necessity implies frequent, perhaps nightly revisits to the same and neighboring fields (since the objects move). Thus a three-nightly-visit cadence permits the near-Sun twilight program to be swapped in relatively freely with other possible twilight programs distant from low solar elongations. This will also relatively gracefully allow for stretches of poor weather. Obviously a near-Sun program should not be scheduled when a crescent moon is present in the field of regard.

\subsection{Filter choice}

No constraints, but generally the redder, broader filters will be preferred.  We anticipate that LSST considerations outside this near-Sun survey will control the selection of filters available during twilight. Due to the short duration of each twilight period, we expect a single filter will be available each evening or morning, and this is sufficient for our key science goals which focuses on detection, orbit characterization, and activity searches. We note, the search for activity in these Solar System bodies is possible in any filter band.

\subsection{Exposure constraints}

Maximizing the area of sky covered during each twilight will generally be the best strategy for a discovery-oriented mini-survey, arguing for a short, constant exposure time to permit the greatest number of exposures during the brief twilight period. Also, tracklet detectability will be limited by the shortest exposure in any event. The maximum exposure time per visit will be controlled by LSST's practical definition of "twilight". The minimum exposure time is one second or perhaps a few seconds to keep the slew, shutter, and readout overheads manageable. Generally a NEO search strategy benefits from going wide versus going deep, but these trade-offs could be modeled with the application of a sky brightness model and other external survey constraints.

\subsection{Other constraints}

None.

\subsection{Estimated time requirement}

Since this is a twilight survey, time is estimated most usefully as a simple count of either evening or morning twilight periods that are allocated to this project. Throughout the duration of LSST survey operations, NEOs and other moving objects will enter the field of regard of the L4/L5 picket line pattern of this survey. The near-Sun survey can productively use as many twilight periods as may be allocated. If these are shared with other non near-Sun science programs, we would request initial scheduling early during LSST operations.

\begin{table}[ht]
    \centering
    \begin{tabular}{l|l|l|l}
        \toprule
        Properties & Importance \hspace{.3in} \\
        \midrule
        Image quality & 3    \\
        Sky brightness & 3 \\
        Individual image depth & 3  \\
        Co-added image depth & 3  \\
        Number of exposures in a visit & 3  \\
        Number of visits (in a night) & 1  \\ 
        Total number of visits & 3  \\
        Time between visits (in a night) & 1 \\
        Time between visits (between nights) & 3  \\
        Long-term gaps between visits & 3 \\
        Other (please add other constraints as needed) & 3 \\
        \bottomrule
    \end{tabular}
    \caption{{\bf Constraint Rankings:} Summary of the relative importance of various survey strategy constraints.}
        \label{tab:obs_constraints}
\end{table}

\subsection{Technical trades}
Since twilight periods are short, it is likely that a particular evening or morning twilight will be allocated to a single program. Solar System science as described in section 2 will require near-Sun targeting, for instance, and no attempt would be made to detect moving objects for these science cases from near-zenith twilight fields. On the other hand, the Catalina Real-Time Survey (CRTS\footnote{\emph{http://crts.caltech.edu/pub.html}}) demonstrated that NEO-optimized survey data can be used to harvest stationary celestial transients of diverse types, see question 5 below.\\ 

    {\it 1. What is the effect of a trade-off between your requested survey footprint (area) and requested co-added depth or number of visits?}\\
    
Moving objects will be detected in single visit observations. The co-added depth is not relevant for our science case but the number of visits is. Multiple visits to the field during twilight will be used to identify moving Solar System objects. If there are more visits per field in a given night, then there would be less sky-coverage. We have selected the number of repeats to each target field to balance this tension. \\
    
    {\it 2. If not requesting a specific timing of visits, what is the effect of a trade-off between the uniformity of observations and the frequency of observations in time? e.g. a `rolling cadence' increases the frequency of visits during a short time period at the cost of fewer visits the rest of the time, making the overall sampling less uniform.}\\
    
Not applicable.\\
    
    {\it 3. What is the effect of a trade-off on the exposure time and number of visits (e.g. increasing the individual image depth but decreasing the overall number of visits)?}\\
    
The twilight nature of this mini-survey constrains any such trade-offs. Exposures must be kept short due to the high and varying sky brightness. The limited overall period of each twilight session then controls the number of fields that can be visited. In addition the moving object detection requires multiple revisits to each field in a short period of time.\\
    
    {\it 4. What is the effect of a trade-off between uniformity in number of visits and co-added depth? Is there any benefit to real-time exposure time optimization to obtain nearly constant single-visit limiting depth?}\\
    
Not applicable.\\
    
    {\it 5. Are there any other potential trade-offs to consider when attempting to balance this proposal with others which may have similar but slightly different requests?}\\

The essence of LSST is to serve multiple science programs with a single data set. This has been demonstrated successfully in the past with the Catalina Real-Time Survey (CRTS) harvesting celestial transients from data collected by the Catalina Sky Survey to look for Near-Earth Objects. The normal LSST workflow should be able to detect transients in the near-Sun field of regard of this twilight survey.

\section{Performance Evaluation}
%\begin{footnotesize}
%{\it Please describe how to evaluate the performance of a given survey in achieving your desired
%science goals, ideally as a heuristic tied directly to the observing strategy (e.g. number of visits obtained
%within a window of time with a specified set of filters) with a clear link to the resulting effect on science.
%More complex metrics which more directly evaluate science output (e.g. number of eclipsing binaries successfully
%identified as a result of a given survey) are also encouraged, preferably as a secondary metric.
%If possible, provide threshold values for these metrics at which point your proposed science would be unsuccessful 
%and where it reaches an ideal goal, or explain why this is not possible to quantify. While not necessary, 
%if you have already transformed this into a MAF metric, please add a link to the code (or a PR to 
%\href{https://github.com/lsst-nonproject/sims_maf_contrib}{sims\_maf\_contrib}) in addition to the text description. %(Limit: 2 pages).}
%\end{footnotesize}

Our main science goals, outlined in the Scientific Motivation, are primarily discovery-based. Our aim is to increase the number of detections in the key small body populations most benefited by a near-Sun/twilight observations (including Near Earth Objects, Interior Earth Objects, Potentially Hazardous Asteroids, Earth Trojans, near-Sun asteroids, sun-grazing comets, and dormant comets). Any observing time that LSST can devote to such a survey in morning and evening twilight will contribute towards this goal. 

To assess the success of a twilight survey design, we propose two discovery metrics: (1) the standard MOPS metric of 3 tracklets obtained on separate nights, distributed over a maximum time period of 15 nights; and, (2) a more typical NEO-survey metric of 4 detections from a single night.  The choice of metric will depend on the style of survey: (1) would be best for a twilight survey with limited nightly field repeats, (2) when several repeats are employed (see Section 3.2). Our survey design is flexible, and thus we can make use of any twilight time that is available.  These success metrics can be tested using the number of NEA detections using the orbital and size distribution from \cite{Granvik2016}. 

Sun-grazing comets have peculiar orbital parameters when compared with NEOs, so we address them specifically here.  Our goal is to discover a sufficient population of the largest Kreutz objects in order to estimate their mass-loss rates at 1~au, or otherwise produce strict upper-limits (cf.~\cite{Ye2014}).  We expect that other sun-grazing populations may be observed by LSST, but use the Kreutz group as a stand-in for assessing survey design.  We propose that a few discoveries per Kreutz observing season (Nov--Jan) is sufficient to probe the range and typical behavior of these objects.  %Here, a potential discovery is an observation of the inbound path of a sun-grazing comet that is later seen by a solar observing spacecraft.
Knight et al.~\cite{Knight2010} estimated the cumulative size distribution of Kreutz sun-grazers, and found them to follow $R^{-2.2}$ from 35 to 5~m in radius, or absolute magnitude $24.9 < H < 29.1$ (largest was $H=24.1$ or 50~m).  %These coarse size estimates have a factor of 2 uncertainty (1.5~mag).
Within the orbital parameters of the Kreutz group, $H=24$~mag corresponds to $r\sim26\pm1.5$~mag for LSST, i.e., inactive nuclei will not be observed.  However, if these comets have a similar activity level (mass-loss rate per unit surface area) as other comets at 1~au, then we expect $r\sim$23~mag (typical twilight survey depth) for 20-m objects.  %, corresponding to a water production rate of $\sim10^{25}$ molecules s$^{-1}$.
The rate of discoveries will indicate what level of activity (and therefore near surface water content) these comets have -- if it is near 100\% then LSST will see such a comet every few days, while, on the other hand, no discoveries over the full survey will imply active areas under 1\% of the surface of these small comets. 

%{\bf CS:} need to be careful here, I think we only find small Kreutz things if they are active, they will be too small if inactive, even with LSST? will check numbers later.

\section{Special Data Processing}
The near-Sun twilight survey is designed to minimize special data processing. Images will be standard, but shorter duration, comparable to the WFD survey. Data will be processed using the standard pipeline, transient detection, and identification of moving objects and their linking to catalog objects and earlier LSST tracklets via MOPS. The project already plans to identify tracklets with more than two observations per night, and is considering whether to report high-confidence single-night tracklets immediately to the MPC. If the project is not able to do so, utility applications may be developed by the community in support of this survey (e.g., for efficient human review, validation of moving object candidates, and submission to the MPC). Such community-driven tool could be equally applicable to WFD and other mini-surveys.

\section{Acknowledgements}
 The authors thank the  Large Synoptic Survey Telescope (LSST) Project Science Team and the LSST Corporation for their support of LSST Solar System Science Collaboration's (SSSC) efforts. This work was supported in part by a LSST Corporation Enabling Science grant. The authors also thank the B612 Foundation, AURA, and the Simons Foundation for their support of workshops, hackathons, and sprints that lead to the development of this white paper. Elements of this work were enabled by the Solar System JupyterHub service at the University of Washington's DIRAC Institute (\url{http://dirac.astro.washington.edu}). This white paper has made use of NASA's Astrophysics Data System Bibliographic Services.
 
%\section{References}

%\bibliography{references} 

\end{document}